# The Trailer of Blockchain Governance Game

SONG-KYOO KIM

**ABSTRACT**

This paper deals with the design of the secure blockchain network framework to prevent damages from an attacker. The decentralized network design called the *Blockchain Governance Game* is a new hybrid theoretical model and it provides the stochastic game framework to find best strategies towards preparation for preventing a network malfunction by an attacker. Analytically tractable results are obtained by using the fluctuation theory and the mixed strategy game theory. These results enable to predict the moment for operations and deliver the optimal portion of backup nodes to protect the blockchain network. This research helps for whom considers the initial coin offering or launching new blockchain based services with enhancing the security features.

## 1. INTRODUCTION

The cryptocurrencies (coins and tokens) are the collection of concepts and technologies that form the basis of a digital money ecosystem (Antonopoulos, 2017). Units of currency called coins (or tokens) are used to store and transmit value among participants in the blockchain network. The Bitcoin is the first cryptocurrency based on the blockchain which has been proposed by Satoshi (Nakamoto, 2009) who is a founder of the Bitcoin. A blockchain is a growing list of records which are linked using cryptography. After establishing the Bitcoin, many cryptocurrencies including the Ethereum (Wood, 2018) have been developed based on the Blockchain technology (Kim, 2018b). Although blockchain records are not unalterable, blockchains could be considered as highly secure by design and exemplify a distributed computing system (Luu, Teutsch and et. al., 2015). One of strength of blockchain is the decentralized peer-to-peer network which eliminates a number of security risks that come with data being held centrally. A vital concern in the Bitcoin is the prevention of unauthorized double-spending. An unknown user by using large amounts of hash power could try to use the 51 percent attack to perform double spending to steal money from exchanges (Garay, Kiayias and et. al., 2015). Decentralized consensus has been claimed with a blockchain and we have observed the intensive attacks which are dedicated for decentralized networks (i.e., Blockchain networks). The 51 percent attack is one of typical attacks by generating blocks with false information of transactions because of this genie strength of the Blockchain (Beikverdi and Song, 2015; Yli-Huumo, Ko and et. al., 2016). The Ethereum has some improvements in terms of managing the authorization of funds and authenticating external entities (Hirai, 2017). But current blockchain based cryptocurrencies including the Ethereum are not fully safe from the 51 percent attack especially related with the mining computation. Because of this reason, a private blockchain has been proposed for business or government uses (Kim, 2018a; Narayanan and Clar, 2017) and this type of Blockchain networks can be considered as a middle-ground for companies and governments (Weiss and Corsi, 2018). Although many peoples are interested in the private Blockchain technology in general but they are not comfortable with a level of control compare to the offered control level from a public decentralized network (Kim, 2018a). In addition, even the control levels for a private blockchain network should be minimal to keep the strengths of a decentralized network for avoiding all security matters what atypical centralized network has.

Recent researchers have improved the securities in the protocol levels and some researches have proposed the new protocol to prevent the 51 percent attack (Decker and Wattenhofer, 2013; Eyal and Sirer, 2014; Garay, Kiayias and Leonardos, 2015; Luu and et. al., 2015). The blockchain protocol requires to solve a POW (Proof-Of-Work), which essentially amounts to brute-forcing a hash inequality based on SHA-256 to create new blockchain blocks (Garay, Kiayias and Leonardos, 2015) and hashcash which has been originally used for preventing the denial of service attack (DoS) has been applied for the POW algorithm in the Blockchain protocol (Back, 2002) and the POW algorithm also has been improved for protecting the decentralized network from attacks (Laurie and Clayton, 2004; Poelstra, 2015). These protocol improvements might prevent the 51 percent attack but most solutions are limited because the implementations are robust by arbitrarily choosing the boundaries (Eyal and Sirer, 2014; Garay, Kiayias and Leonardos, 2015). Unfortunately, a thorough analysis for establishing the exact security properties (i.e., the optimized boundaries) of the Blockchain system has not been established yet.

The *Blockchain Governance Game* is newly proposed in the paper. The *Blockchain Governance Game* (BGG) is the stochastic game model with the fluctuation and the mixed strategy game for analyzing the network to provide the decision making moment for taking preliminary security actions before attacks. The model is targeted to prevent blockchain based attacks (i.e., the 51 percent attack) and keeps the network decentralized. Atypical case that an attacker to trying to build an alternative blockchain (blockchain forks) faster than regular miners (Nakamoto, 2009) is considered in BGG. The defender (or the controller) only manages the small percentage of nodes which are released prior the attack is happened. The results are given as joint functionals between two players of the predicted time of the first observed threshold which is crossing the half of the total nodes (i.e., 51 percents) along with values of each component upon this time.

## 2. Stochastic Model For Blockchain Network

### 2.1 Basic Stochastic Model

The antagonistic game of two players (called "A" and "H") are introduced to describe the blockchain network between a defender and an attacker. Both players compete to build the blocks either for honest or false ones. Let ($\Omega$, $\mathcal{F}(\Omega)$,$P$) be probability space $\mathcal{F}_A$, $\mathcal{F}_H$, $\mathcal{F}_\tau \subseteq \mathcal{F}(\Omega)$ be independent $\sigma$-subalgebras. Suppose:

$$\mathcal{A} := \sum_{k\geq 0} X_k \varepsilon_{s_k},\ s_0(=0) < s_1 < s_2 < \cdots, \text{ a.s.} \tag{2.1}$$

$$\mathcal{H} := \sum_{j\geq 0} Y_j \varepsilon_{t_j},\ t_0(=0) < t_1 < t_2 < \cdots, \text{ a.s.} \tag{2.2}$$

are $\mathcal{F}_A$-measurable and $\mathcal{F}_H$-measurable marked Poisson processes ($\varepsilon_a$ is a point mass at $a$) with respective intensities $\lambda_A$ and $\lambda_H$ and point independent marking. These two values are related with the computing performance for generating blocks for attackers and honest nodes in the blockchain network. They will represent the actions of player A (an attacker) and H (an honest node). Player A builds the blocks with false transactions (e.g., double spend) at times $s_1,\ s_2,\ldots$ and sustain respective build the blocks of magnitudes $X_1,\ X_{2,\ldots}$ formalized by the process $\mathcal{A}$. The building blocks to player H are described by the process $\mathcal{H}$ similarly. Player H will generate the blocks which contain the correct transactions. Both players races to build their blocks (either honest or false). The processes $\mathcal{A}$ and $\mathcal{H}$ are specified by their transforms

$$\mathbb{E}\left[g^{\mathcal{A}(s)}\right] = e^{\lambda_A(s)(g-1)},\ \mathbb{E}\left[z^{\mathcal{H}(t)}\right] = e^{\lambda_H(t)(z-1)}. \tag{2.3}$$

The game is observed at random times in accordance with the point process which is equivalent with the duration of the PoW (Proof-of-Work) completion (around 10 minutes in the Bitcoin) in the blockchain network (Kim, 2018a):

$$\mathcal{T} := \sum_{i\geq 0} \varepsilon_{\tau_i},\ \tau_0(>0)),\tau_1,\ldots, \tag{2.4}$$

which is assumed to be delayed renewal process. If

$$(A(t), H(t)) := \mathcal{A} \otimes \mathcal{H}([0,\tau_k]),\ k = 0,1,\ldots, \tag{2.5}$$

forms an observation process upon $\mathcal{A} \otimes \mathcal{H}$ embedded over $\mathcal{T}$, with respective increments

$$(X_k, Y_k) := \mathcal{A} \otimes \mathcal{H}([\tau_{k-1},\tau_k]),\ k = 1,2,\ldots, \tag{2.6}$$

and

$$X_0 = A_0,\ Y_0 = H_0. \tag{2.7}$$

The observation process could be formalized as

$$\mathcal{A}_\tau \otimes \mathcal{H}_\tau := \sum_{k\geq 0} (X_k, Y_k)\varepsilon_{\tau_k}, \tag{2.8}$$

where

$$\mathcal{A}_\tau = \sum_{i \geq 0} X_i \varepsilon_{\tau_i},\ \mathcal{H}_\tau = \sum_{i \geq 0} Y_i \varepsilon_{\tau_i}, \tag{2.9}$$

and it is with position dependent marking and with $X_k$ and $Y_k$ being dependent with the notation

$$\Delta_k := \tau_k - \tau_{k-1},\ k = 0, 1, \ldots,\ \tau_{-1} = 0, \tag{2.10}$$

and

$$\gamma(g, z) = \mathbb{E}\big[g^{X_k} \cdot z^{Y_k}\big],\ g > 0,\ z > 0. \tag{2.11}$$

By using the double expectation,

$$\gamma(g, z) = \delta(\lambda_A(1-g) + \lambda_H(1-z)), \tag{2.12}$$

and

$$\gamma_0(g, z) = \mathbb{E}\big[g^{A_0} z^{H_0}\big] = \delta_0(\lambda_A(1-g) + \lambda_H(1-z)), \tag{2.13}$$

where

$$\delta(\theta) = \mathbb{E}\big[e^{-\theta \Delta_1}\big],\ \delta_0(\theta) = \mathbb{E}\big[e^{-\theta \tau_0}\big], \tag{2.14}$$

are the magical transform of increments $\Delta_1, \Delta_2, \ldots$. The game in this case is a stochastic process $\mathcal{A}_\tau \otimes \mathcal{H}_\tau$ describing the evolution of a conflict between players A and H known to an observation process $\mathcal{T}$=$\{\tau_0, \tau_1, \ldots\}$. The game is over when on the $k$th observation epoch $\tau_k$, the collateral building blocks to player A exceeds more than the half of the total nodes $M$. To further formalize the game, the *exit index* is introduced:

$$\nu := \mathit{inf}\big\{k : A_k = A_0 + X_1 + \cdots + X_k \geq \big(\tfrac{M}{2}\big)\big\}, \tag{2.15}$$

$$\mu := \mathit{inf}\big\{j : H_j = H_0 + Y_1 + \cdots + Y_j \geq \big(\tfrac{M}{2}\big)\big\}. \tag{2.15a}$$

Since, an attacker is win at time $\tau_\nu$, otherwise an honest node generates the correct blocks. We shall be targeting the confined game in the view point of player A. The first passage time $\tau_\nu$ is the associated exit time from the confined game and the formula (2.6) will be modified as

$$\overline{\mathcal{A}_\tau} \otimes \overline{\mathcal{H}_\tau} := \sum_{k \geq 0}^{\nu} (X_k, Y_k) \varepsilon_{\tau_k} \tag{2.16}$$

which the path of the game from $\mathcal{F}(\Omega) \cap \{\nu < \mu\}$, which gives an exact definition of the model observed until $\tau_\nu$. The joint functional of the blockchain network model is as follows:

$$\begin{aligned} \Phi_{\lceil \frac{M}{2} \rceil} &= \Phi_{\lceil \frac{M}{2} \rceil}(\xi, g_0, g_1, z_0, z_1) \\ &= \mathbb{E}\Big[\xi^\nu \cdot g_0^{A_{\nu-1}} \cdot g_1^{A_\nu} \cdot z_0^{H_{\nu-1}} \cdot z_1^{H_\nu} \mathbf{1}_{\{\nu < \mu\}}\Big], \end{aligned} \tag{2.17}$$

where $M$ indicates the total number of nodes (or ledgers) in the blockchain network. This functional will represent the status of attackers and honest nodes upon the exit time $\tau_\nu$. The latter is of particular interest, we are interested in not only the prediction of catching up the blocks by attackers but also one observation prior to this. The Theorem 1 (BGG-1) establishes an explicit formula $\Phi_{\frac{M}{2}}$ from (2.11)-(2.13). The first exceed model by Dshahalow (1995) has been adopted and its operators are defined as follows:

$$\mathcal{D}_{(x,y)}\left[f(x,y)\right](u,v) := (1-u)(1-v)\sum_{x\geq 0}\sum_{y\geq 0} f(x,y)u^x v^y, \tag{2.18}$$

then

$$f(x,y) = \mathfrak{D}_{(u,v)}^{(x,y)}\left[\mathcal{D}_{(x,y)}\left\{f(x,y)\right\}\right], \tag{2.18a}$$

where $\{f(x,y)\}$ is a sequence, with the inverse

$$\mathfrak{D}_{(u,v)}^{(m,n)}(\,\bullet\,) = \begin{cases} \left(\frac{1}{m!\cdot n!}\right)\lim\limits_{(u,v)\to 0}\frac{\partial^m\partial^n}{\partial u^m \partial v^n}\frac{1}{(1-u)(1-v)}(\,\bullet\,), & m\geq 0, n\geq 0, \\ 0, & \textit{otherwise}. \end{cases} \tag{2.19}$$

**Theorem 1 (BGG-1):** the functional $\Phi_{\frac{M}{2}}$ of the process of (2.17) satisfies following expression:

$$\Phi_{\lceil\frac{M}{2}\rceil} = \mathfrak{D}_{(u,v)}^{(\lceil\frac{M}{2}\rceil,\lceil\frac{M}{2}\rceil)}\left[\Gamma_0^1 - \Gamma_0 + \frac{\xi\cdot\gamma_0}{1-\xi\gamma}\left(\Gamma^1 - \Gamma\right)\right]. \tag{2.20}$$

From (2.17) and (2.35), we can find the PGFs (probability generating functions) of $A_{\nu-1}$ (and $A_\nu$) and the *exit index*:

$$\mathbb{E}[\xi^\nu] = \Phi_{\lceil\frac{M}{2}\rceil}(\xi,1,1,1,1), \tag{2.36}$$

$$\mathbb{E}\left[g_0^{A_{\nu-1}}\right] = \Phi_{\lceil\frac{M}{2}\rceil}(1,g_0,1,1,1), \tag{2.37}$$

$$\mathbb{E}\left[g_1^{A_\nu}\right] = \Phi_{\lceil\frac{M}{2}\rceil}(1,1,g_1,1,1). \tag{2.38}$$

The moment of making a decision $\tau_{\nu-1}$ could be found from (2.4), (2.10) and (2.36):

$$\mathbb{E}[\nu] = \tfrac{\partial}{\partial\xi}\Phi_{\lceil\frac{M}{2}\rceil}(\xi,1,1,1,1)\Big|_{\xi=1}, \tag{2.39}$$

$$\mathbb{E}[\tau_{\nu-1}] = \mathbb{E}[\tau_0] + \mathbb{E}[\Delta_1](\mathbb{E}[\nu]-1). \tag{2.40}$$

# 3. Blockchain Governance Mixed Game Strategy

## 3.1 Preliminaries

Let us consider a two-person mixed strategy game, and player H (i.e., a defender) is the person who has two strategies at the observation moment, one step before attackers complete to generate alternative chains with dishonest transactions (double spending). Player H has the following strategies: (1) *DoNothing* – doing nothing, which implicates that the blockchain networks are running as usual, and (2) *Action* – taking the preliminary action for avoiding attacks by adding honest nodes. In the view of player A (an attacker), he might succeed to catch the blocks or fail to catch (i.e., the Blockchain network has been defended). Therefore, the responses of player A would be either "Not burst" or "Burst." Let us assume that the cost for reserving the additional honest nodes is $c_\alpha$ where $\alpha$ is the portion to reserve the blocks for defending the Blockchain network. The token provider might reserve the certain portion of nodes for protecting the values and the network. If the attacks succeed to generate alternative blocks, the network bursts and the whole value of the tokens (or coins) $B$ will be lost and this value might be equivalent with the value by ICO (Initial Coin Offering). It still has the chance to burst although the defender (or the provider) adds the honest nodes before catching blocks by the attacker. In this case, the cost will be not only the token value but also the reservation cost for additional honest nodes. The normal form of games is as follows:

$$\begin{aligned}
&\text{. Players:} && \boldsymbol{N} = \{A, H\}, \\
&\text{. Strategy sets:} && \\
& && \boldsymbol{s_A} = \{\text{"}\textit{NotBurst}\text{", "}\textit{Burst}\text{"}\}, \\
& && \boldsymbol{s_H} = \{\text{"}\textit{DoNothing}\text{", "}\textit{Action}\text{"}\}.
\end{aligned} \tag{3.1}$$

Based on the above conditions, the general cost matrix at the prior time to be burst $\tau_{\nu-1}$could be composed as follows:

| | *NotBurst* $(1 - q(s_H))$ | *Burst* $(q(s_H))$ |
|---|---|---|
| *DoNothing* | $0$ | $B$ |
| *Action* | $c_\alpha$ | $c_\alpha + B$ |

**Table 1.** Cost matrix

where $q(s_H)$ is the probability of bursting blockchain network (i.e., an attacker wins the game) and it depends on the strategic decision of player H (a defender):

$$q(s_H) = \begin{cases} \mathbb{E}\left[\mathbf{1}_{\{A_\nu \geq \frac{M}{2}\}}\right], & s_H = \{\textit{DoNothing}\}, \\ \mathbb{E}\left[\mathbf{1}_{\left\{A_\nu \geq \frac{M(1+\alpha)}{2}\right\}}\right], & s_H = \{\textit{Action}\}. \end{cases} \tag{3.2}$$

It is noted that the cost for the reserved nodes (i.e., "Action" strategy of player H) should be smaller than the other strategy. Otherwise, player H does not have to spend the cost of the governance protection. The portion of reserved nodes for protecting a blockchain network $\alpha$ depends on the cost function and the optimal portion for the blockchain governance $\alpha^*$ could be found as follows:

$$\alpha^* = \textit{inf}\left\{\alpha \geq 0 : \mathfrak{C}_{\textit{NoA}}\left(q^0\right) \geq \mathfrak{C}_{\textit{Act}}(\alpha)\right\}, \tag{3.3}$$

where (at the moment $\tau_{\nu-1}$),

$$\mathfrak{C}_{\textit{NoA}}\left(q^0\right) = B \cdot q^0, \tag{3.4}$$

$$\mathfrak{C}_{\textit{Act}}(\alpha) = c_\alpha\left(1 - q_\alpha^1\right) + (c_\alpha + B)q_\alpha^1, \tag{3.5}$$

$$q^0 = \mathbb{E}\left[\mathbf{1}_{\{A_\nu \geq \frac{M}{2}\}}\right], q_\alpha^1 = \mathbb{E}\left[\mathbf{1}_{\left\{A_\nu \geq \frac{M(1+\alpha)}{2}\right\}}\right]. \tag{3.6}$$

### 3.2 Blockchain Governance Game

We would like to design the enhanced blockchain network governance that can take the action at the decision making moment $\tau_{\nu-1}$. The governance in the blockchain is followed by the decision making parameter. It also means that we will not take any action until the time $\tau_{\nu-1}$ and it still have the chance that all nodes are governed by an attacker if the attacker catches more than the half of nodes at $\tau_{\nu-1}$ (i.e., $\left\{A_{\nu-1} \geq \frac{M}{2}\right\}$. If the attacker catches the less than half of all nodes at $\tau_{\nu-1}$ (i.e., $\left\{A_{\nu-1} < \frac{M}{2}\right\}$, then the defender could take the action to avoid the attack at $\tau_\nu$. The total cost for developing the enhanced blockchain network is as follows:

$$\begin{aligned}\mathfrak{C}(q^0,\alpha)_{Total} &= \mathbb{E}\Big[\mathfrak{C}_{Act}(\alpha)\cdot \mathbf{1}_{\{A_{\nu-1}<\frac{M}{2}\}} + \mathfrak{C}_{NoA}(q^0)\cdot \mathbf{1}_{\{A_{\nu-1}\geq \frac{M}{2}\}}\Big] \\ &= \{c_\alpha(1-q_\alpha^1)+(c_\alpha+B)q_\alpha^1\}p_{A_{-1}} + B\cdot q^0(1-p_{A_{-1}})\end{aligned} \tag{3.7}$$

where

$$p_{A_{-1}} = \boldsymbol{P}\Big\{A_{\nu-1} < \frac{M}{2}\Big\} = \sum_{k=0}^{\lfloor \frac{M}{2}\rfloor} \boldsymbol{P}\{A_{\nu-1}=k\}. \tag{3.8}$$

Because $\varPhi_{\lceil \frac{M}{2}\rceil}(1,g_0,1,1,1)$ from (2.17) is the probability generating function (PGF) of $A_{\nu-1}$, the probability mass could be found as follows:

$$\boldsymbol{P}\{A_{\nu-1}=k\} = \lim_{g_0\to 0}\frac{1}{k!}\frac{\partial^k}{\partial g_0^k}\varPhi_{\lceil \frac{M}{2}\rceil}(1,g_0,1,1,1),\ k=0,\dots. \tag{3.9}$$

## 4. Special Case: Memoryless Observation Process

It is assumed that the observation process has the memoryless properties which might be a special condition but very practical for actual implementation on a blockchain governance. It implies that the defender (or a service provider) does not spend additional cost of storing the past information. To build the cost function of the blockchain governance, we can find explicit solutions of $q^0$, $p_{A_{-1}}$ and the moment of the decision making after finding the first exceed index $\mathbb{E}[\nu]$, the probability (generating function) of the number of blocks at the moment $\tau_\nu$ $\left(\mathbb{E}\left[g_1^{A_\nu}\right]\right)$ and $\tau_{\nu-1}$ $\left(\mathbb{E}\left[g_0^{A_{\nu-1}}\right]\right)$. Recall from (2.18), the operator is defined as follows:

$$G(u) = \mathcal{D}_x\,[f(x)](u) := (1-u)\sum_{x\geq 0} f(x)u^x, \tag{2.18}$$

$$\begin{aligned}\mathcal{D}_{(x,y)}\,[f_1(x)f_2(y)](u,v) &:= (1-u)(1-v)\sum_{x\geq 0}\sum_{y\geq 0} f_1(x)f_2(y)u^x v^y \\ &= \mathcal{D}_x\,[f_1(x)]\mathcal{D}_y\,[f_2(y)],\end{aligned} \tag{4.1}$$

then

$$f(x,y) = \mathfrak{D}_{(u,v)}^{(x,y)}\big[\mathcal{D}_{(x,y)}\,\{f(x,y)\}\big], \tag{4.2}$$

$$f_1(x)f_2(y) = \mathfrak{D}_u^x[\mathcal{D}_x\,\{f_1(x)\}]\mathfrak{D}_v^y[\mathcal{D}_y\,\{f_2(y)\}], \tag{4.3}$$

where $\{f(x),\ (f_1(x)f_2(y))\}$ are a sequence, with the inverse (2.19) and

$$\mathfrak{D}_u^m(\,\bullet\,) = \begin{cases} \frac{1}{m!}\lim\limits_{u\to 0}\frac{\partial^m}{\partial u^m}\frac{1}{(1-u)}(\,\bullet\,), & m\geq 0, \\ 0, & \textit{otherwise,}\end{cases} \tag{4.4}$$

and

$$\mathfrak{D}^{(m,n)}_{(u,v)}[G_1(u)G_2(v)] = \mathfrak{D}^m_u[G_1(u)]\mathfrak{D}^n_v[G_2(v)]. \tag{4.5}$$

The functional $\mathfrak{D}$ is defined on the space of all analytic functions at 0 and it has the following properties:

(i) $\mathfrak{D}^m_u$ is a linear functional with fixed points at constant functions,

$$\text{(ii) } \mathfrak{D}^m_u \sum_{k=0}^{\infty} a_k u^k = \sum_{k=0}^{m} a_k. \tag{4.6}$$

It is also noted that the formulas (2.11)-(2.14) could be rewritten as follows:

$$\gamma(g,z) = \delta(\lambda_A(1-g)+\lambda_H(1-z)) = \gamma_A(g)\cdot\gamma_H(z), \tag{4.7}$$
$$\gamma_A(g) = \delta(\lambda_A(1-g)), \tag{4.8}$$
$$\gamma_H(z) = \delta(\lambda_H(1-z)), \tag{4.9}$$

and

$$\gamma_0(g,z) = \delta_0(\lambda_A(1-g)+\lambda_H(1-z)) = \gamma^0_A(g)\cdot\gamma^0_H(z), \tag{4.10}$$
$$\gamma^0_A(g) = \mathbb{E}\big[g^{A_0}\big] = \delta_0(\lambda_A(1-g)), \tag{4.11}$$
$$\gamma^0_H(z) = \mathbb{E}\big[z^{H_0}\big] = \delta_0(\lambda_H(1-z)), \tag{4.12}$$

from (2.29)-(2.34),

$$\gamma = \gamma_A\cdot\gamma_H := \gamma_A(g_0g_1u)\gamma_H(z_0z_1v), \tag{4.13}$$
$$\gamma_0 = \gamma^0_A\cdot\gamma^0_H := \gamma^0_A(g_0g_1u)\gamma^0_H(z_0z_1v), \tag{4.14}$$

$$\mathit{\Gamma} := \gamma_A(g_1u)\gamma_H(z_1v), \tag{4.15}$$
$$\mathit{\Gamma}_0 := \gamma^0_A(g_1u)\gamma^0_H(z_1v), \tag{4.16}$$

$$\mathit{\Gamma}^1 := \gamma_A(g_1)\gamma_H(z_1v), \tag{4.17}$$
$$\mathit{\Gamma}^1_0 := \gamma^0_A(g_1)\gamma^0_H(z_1v). \tag{4.18}$$

### 4.1. The Marginal Mean upon $\tau_{\nu-1}$

The marginal mean of $\tau_{\nu-1}$ is the moment of the decision making for taking the prior action to prevent the attack and it could be straight forward once the exit index is found. The *exit index* could be found from (2.29)-(2.34), (2.36) and (4.7)-(4.18):

$$\mathbb{E}[\xi^\nu] = \mathit{\Phi}_{\lceil\frac{M}{2}\rceil}(\xi,1,1,1,1) = L^1 + L^2 - L^3 \tag{4.19}$$

where

$$L^1 = \mathfrak{D}^{(\frac{M}{2},\frac{M}{2})}_{(u,v)}[\gamma^0_H(v) - \gamma^0_A(u)\gamma^0_H(v)], \tag{4.20}$$
$$L^2 = \mathfrak{D}^{(\frac{M}{2},\frac{M}{2})}_{(u,v)}\left[\frac{\xi\cdot\gamma^0_A(u)\gamma^0_H(v)\gamma_H(v)}{1-\xi\gamma_A(u)\gamma_H(v)}\right], \tag{4.21}$$
$$L^3 = \mathfrak{D}^{(\frac{M}{2},\frac{M}{2})}_{(u,v)}\left[\frac{\xi\cdot\gamma^0_A(u)\gamma^0_H(v)\gamma_A(u)\gamma_H(v)}{1-\xi\gamma_A(u)\gamma_H(v)}\right]. \tag{4.22}$$

Since, the observation process has the memoryless properties, the process is exponentially distributed and the functionals from (4.7)-(4.12) are as follows:

$$\gamma_A^0(u) = \frac{1}{\left(1+\widetilde{\delta_0}\cdot\lambda_A\right)-\widetilde{\delta_0}\cdot\lambda_A u} = \frac{\beta_A^0}{1-\alpha_A^0\cdot u}, \tag{4.23}$$

$$\gamma_A(u) = \frac{1}{\left(1+\widetilde{\delta}\cdot\lambda_A\right)-\widetilde{\delta}\cdot\lambda_A u} = \frac{\beta_A}{1-\alpha_A\cdot u}, \tag{4.24}$$

$$\gamma_H^0(v) = \frac{1}{\left(1+\widetilde{\delta_0}\cdot\lambda_H\right)-\widetilde{\delta_0}\cdot\lambda_H v} = \frac{\beta_H^0}{1-\alpha_H^0\cdot v}, \tag{4.25}$$

$$\gamma_H(v) = \frac{1}{\left(1+\widetilde{\delta}\cdot\lambda_H\right)-\widetilde{\delta}\cdot\lambda_H v} = \frac{\beta_H}{1-\alpha_H\cdot v}, \tag{4.26}$$

$$\beta_A^0 = \frac{1}{\left(1+\widetilde{\delta_0}\cdot\lambda_A\right)},\ \alpha_A^0 = \frac{\widetilde{\delta_0}\cdot\lambda_A}{\left(1+\widetilde{\delta_0}\cdot\lambda_A\right)}, \tag{4.27}$$

$$\beta_A = \frac{1}{\left(1+\widetilde{\delta}\cdot\lambda_A\right)},\ \alpha_A = \frac{\widetilde{\delta_0}\cdot\lambda_A}{\left(1+\widetilde{\delta}\cdot\lambda_A\right)}, \tag{4.28}$$

$$\beta_H^0 = \frac{1}{\left(1+\widetilde{\delta_0}\cdot\lambda_H\right)},\ \alpha_H^0 = \frac{\widetilde{\delta_0}\cdot\lambda_H}{\left(1+\widetilde{\delta_0}\cdot\lambda_H\right)}, \tag{4.29}$$

$$\beta_H = \frac{1}{\left(1+\widetilde{\delta}\cdot\lambda_H\right)},\ \alpha_H = \frac{\widetilde{\delta_0}\cdot\lambda_H}{\left(1+\widetilde{\delta}\cdot\lambda_H\right)}, \tag{4.30}$$

where

$$\widetilde{\delta_0} = \mathbb{E}[\tau_0],\ \widetilde{\delta} = \mathbb{E}[\Delta_k]. \tag{4.31}$$

$$\mathbb{E}[\xi^\nu] = \beta_H^0\left[\frac{1-(\alpha_H^0)^{\frac{M}{2}+1}}{1-(\alpha_H^0)}\right]\left(1-\beta_A^0\left[\frac{1-(\alpha_A^0)^{\frac{M}{2}+1}}{1-(\alpha_A^0)}\right]\right) \tag{4.42}$$

$$+\sum_{n\geq 1}\left\{(\beta_A^0\beta_H^0\beta_H)(\beta_A\beta_H)^n\Xi_{n-1}^H\left(\tfrac{M}{2}\right)\left\{\Xi_{n-2}^A\left(\tfrac{M}{2}\right)-\beta_A\Xi_{n-1}^A\left(\tfrac{M}{2}\right)\right\}\right\}\xi^n$$

and

$$\mathbb{E}[\nu] = (\beta_A^0\beta_H^0\beta_H)\sum_{n\geq 1}(n)\left[(\beta_A\beta_H)^n\Xi_{n-1}^H\left(\tfrac{M}{2}\right)\left\{\Xi_{n-2}^A\left(\tfrac{M}{2}\right)-\Xi_{n-1}^A\left(\tfrac{M}{2}\right)\beta_A\right\}\right] \tag{4.43}$$

where

$$\Xi_{-1}^A(m) = 0,\ \Xi_{-2}^A(m) = 0,\ \Xi_{-1}^H(m) = 0. \tag{4.44}$$

From (2.40) and (4.42), we also have

$$\mathbb{E}[\tau_{\nu-1}] = \widetilde{\delta_0} - \widetilde{\delta} \tag{4.45}$$

$$+\widetilde{\delta}\,(\beta_A^0\beta_H^0\beta_H)\left\{\sum_{n\geq 2}n\left[(\beta_A\beta_H)^n\Xi_{n-1}^H\left(\tfrac{M}{2}\right)\left\{\Xi_{n-2}^A\left(\tfrac{M}{2}\right)-\Xi_{n-1}^A\left(\tfrac{M}{2}\right)\beta_A\right\}\right]\right\}.$$

### 4.2. The Marginal Transform upon $A_{\nu-1}$

The number of blocks which governed by an attacker at $\tau_{\nu-1}$ is one of vital factors to analyze the network cost. The function $\mathbb{E}\left[g_0^{A_{\nu-1}}\right]$ is the probability generating function (PGF) of the number of attacked block $A_{\nu-1}$ at the prior moment of exceeding more than a half of the total nodes $M$. The probability distribution of $A_{\nu-1}$ could be found from (3.9) after obtaining the PGF. From (2.37),

$$\mathbb{E}\Big[g_0^{A_{\nu-1}}\Big] = \beta_H^0\left[\frac{1-(\alpha_H^0)^{\frac{M}{2}+1}}{1-(\alpha_H^0)}\right]\left(1-\beta_A^0\left[\frac{1-(\alpha_A^0)^{\frac{M}{2}+1}}{1-(\alpha_A^0)}\right]\right) \tag{4.51}$$

$$+\,(\beta_A^0\beta_H^0\beta_H)\sum_{n\geq 0}(\beta_A\beta_H)^n\{\varXi_{n-1}^A\big(\tfrac{M}{2},g_0\big)\cdot\varXi_n^H\big(\tfrac{M}{2}\big)\}$$

$$-\,(\beta_A^0\beta_A\beta_H^0\beta_H)\Big(\tfrac{g_0\cdot\alpha_A^0}{g_0\cdot\alpha_A^0-\alpha_A}\Big)\sum_{n\geq 0}(\beta_A\beta_H)^n\varXi_n^H\big(\tfrac{M}{2}\big)\zeta_{n-1}\big(\tfrac{M}{2},g_0\big)$$

$$+\,(\beta_A^0\beta_A\beta_H^0\beta_H)\Big(\tfrac{\alpha_A}{g_0\cdot\alpha_A^0-\alpha_A}\Big)\sum_{n\geq 0}(\beta_A\beta_H)^n\varXi_n^H\big(\tfrac{M}{2}\big)\varXi_{n-1}^A\big(\tfrac{M}{2},g_0\big).$$

### 4.3. The Marginal Transform upon $A_\nu$

The number of blocks which governed by an attacker at $\tau_\nu$ is another vital factor to analyze the network cost. The function $\mathbb{E}\Big[g_1^{A_\nu}\Big]$ is the probability generating function (PGF) of the number of attacked block $A_\nu$ at the moment of exceeding more than a half of the total nodes $M$. The probability distribution of $A_\nu$ could be found as follows:

$$\boldsymbol{P}\{A_\nu = l\} = \lim_{g_1\to 0}\frac{1}{l!}\frac{\partial^l}{\partial g_1^l}\varPhi_{\lceil\frac{M}{2}\rceil}(1,1,g_1,1,1),\ l=0,\ldots,\left\lceil\frac{M}{2}\right\rceil, \tag{4.52}$$

after obtaining the PGF of $A_\nu$. From (2.38),

$$\mathbb{E}\Big[g_1^{A_\nu}\Big] = \left\{\beta_H^0\left[\frac{1-(\alpha_H^0)^{\frac{M}{2}+1}}{1-(\alpha_H^0)}\right]\right\}\left\{\frac{\beta_A^0(\alpha_A^0\cdot g_1)^{\frac{M}{2}+1}}{1-\alpha_A^0\cdot g_1}\right\} \tag{4.62}$$

$$+\left\{\frac{(\beta_A^0)^2\beta_H^0\beta_H}{1-\alpha_A^0\cdot g_1}\right\}\sum_{n\geq 0}(\beta_A\beta_H)^n\varXi_n^H\big(\tfrac{M}{2}\big)\varXi_{n-1}^A\big(\tfrac{M}{2},g_1\big)$$

$$-\sum(\beta_A^0\beta_A\beta_H^0\beta_H)\sum_{n\geq 0}(\beta_A\beta_H)^n\{\varXi_n^A\big(\tfrac{M}{2},g_1\big)\cdot\varXi_n^H\big(\tfrac{M}{2}\big)\},$$

where

$$\varXi_{-1}^A(k,g_1)=0,\ \varXi_{-1}^H(k)=0. \tag{4.63}$$

### 4.4. Linear Programming Practice

A network security in a private blockchain based service with the token offering is considered in this subsection. Although the Blockchain network is designed to be decentralized, the service provider should have at least enough power to avoid attacks not only from outsiders but also from insiders. The 51 percent attack still could be happened even in the private blockchain networks. As it is mentioned, the strategy for managing the network reliability is for supporting the additional nodes to give the less chance that an attacker catches blocks with false transactions. In the view point of the service planning, this practice is atypical setup of taken offerings. The example in this paper is targeting 100K users, 2M USD total token values and the cost of backup nodes for the governance is 0.5 USD per an honest (backup) nodes (see Table 2):

| Name | Value | Description |
|---|---|---|
| $M$ | 100,000 [User] | Total number of the nodes in the network |
| $B$ | 2,000,000 [USD] | (Target) total value of tokens (or coins) offered by ICO |
| $c_\alpha$ | $= 0.5\alpha \cdot M$[USD] | Cost for reserving additional nodes to avoid attacks (50 C/Node) |
| $\mathbb{E}[A_0]$ | 10 [Blocks] | Total number of blocks that changed by an attacker at $\tau_0(=0)$ |
| $\mathbb{E}[H_0]$ | 150 [Blocks] | Total number of blocks that changed by an honest node at $\tau_0(=0)$ |

**Table 2.** Initial conditions for the cost function

Since, the model of the *Blockchain Governance Game* has been analytically solved, the values for the cost function and the calculations of the probability distributions are straight forward (see Table 3).

| Name | Formula | Description |
|---|---|---|
| $q^0$ | (2.38), (3.6) | Probability that an attacker catches the blocks more than a half at $\tau_\nu$ |
| $q^1_\alpha$ | (2.38), (3.6) | Probability that an attacker catches after adding reserved nodes |
| $\alpha$ | $-$ | The portion of additional nodes for blockchain protection |
| $\alpha^*$ | (3.3) | The portion of the reserved nodes for minimizing the cost |
| $p_{A_{-1}}$ | (2.37), (3.8) | The probability that an attacker is not succeed at $\tau_{\nu-1}$ |
| $\mathfrak{C}(\alpha)_{\mathit{Total}}$ | (3.7) | The total cost function for the enhanced blockchain network |

**Table 3.** Calculated values from the equations

Let us assume some of values including $q^0$, $p_{A-1}$and $q^1_\alpha$ are already obtained after the calculations. It is noted that the purpose of this section only demonstrates how to structure the Linear Programing (LP) model to optimize the cost. These values are directly applied into the optimization model and described by the following LP model:

*Objective* (3.7)

$$\min U = \mathfrak{C}(\alpha)_{\mathit{Total}} \tag{4.64}$$

*Subject to* (3.3)

$$\alpha \geq \frac{c_\alpha}{B \cdot q^0 - c_\alpha}. \tag{4.65}$$

The total cost $\mathfrak{C}(\alpha)_{\mathit{Total}}$ could be minimized by given $\alpha$. As it is mentioned in Table 3, the parameter $\alpha$ is the portion that a defender (a service provider) reserves nodes to protect the network from an attacker. The below illustration (Fig. 1) is atypical graph of an optimal result by using the *Blockchain Governance Game* based on the given conditions (see Table 2). For this example, the optimal cost is 59.6K USD when the defender reserve the 9.5% of total nodes which is 9,500 reserved nodes for managing the risk from attacks.

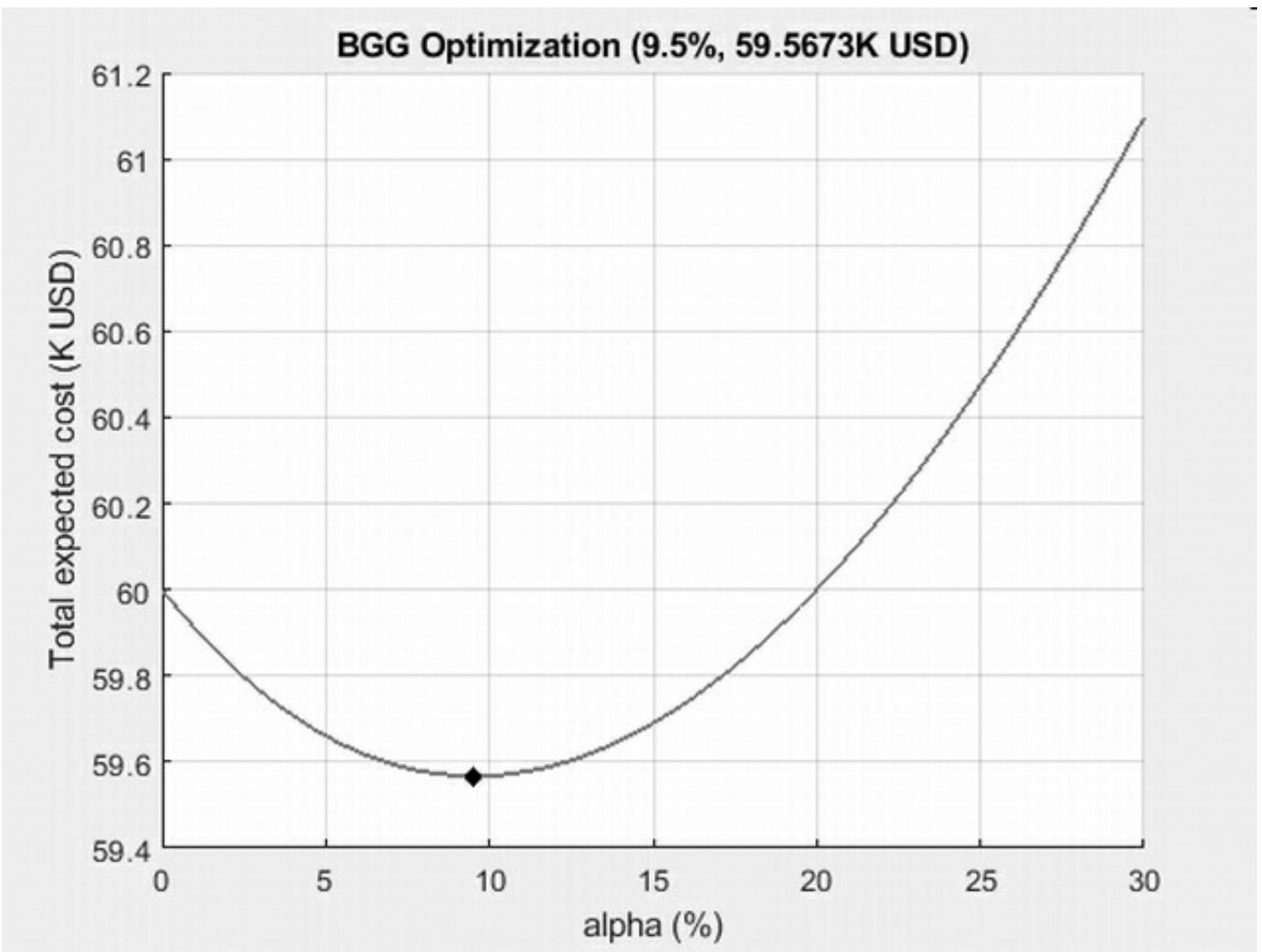

**Fig. 1.** Optimization Example for the Blockchain Governance Game

The moment of releasing the additional nodes will be the time $\tau_{\nu-1}$ when is one step prior than the time when an attacker catches more than half of the whole blocks (i.e., $\tau_{\nu}$).

## 5. Conclusion

The objective of this paper is establishing the theoretical framework of the *Blockchain Governance Game* with the explicit equations for developing the Blockchain network security for avoiding the attacks of the decentralized networks. The core parts of the research including the proof of the Theorem 1 (BGG-1), the analytic functionals for the decision making parameters and the special case are fully deployed in this paper. This research will be helpful for whom considers the Initial Coin Offering (ICO) or launching a new blockchain based service with the network security enhancement. The *Blockchain Governance Game* could be enhanced by gathering all related data from real networks and setting up the initial parameters based on the real measured data is one of ways to enhance the BGG as a future research topic.